
\input phyzzx

\pagenumber=-1
\pubnum={OU-HET-194}
\titlepage
\title{A New Solution of the Solar Neutrino Flux}
\author{ Kazutoshi Ohta and Eiichi Takasugi}
\address{\sl Department of Physics\break
             ~Osaka University, 1-16 Toyonaka, Osaka 560, ~JAPAN
             }
\abstract

We report a new solution to explain the observed deficit of
the solar neutrino flux by Homestake, Kamiokande II and III,
GALLEX and SAGE experiments. We use the matter mixing and
the helicity oscillation in the twisting magnetic fields in
the sun. Our model predicts the short (seasonal) and
long (11 years) time variations of the solar neutrino flux.
Three kinds of data observed by Homestake, Kamiokande, GALLEX and SAGE
detectors are reproduced well if the mixing angle and the squared
mass difference are in the small area around
$\sin^2 2\theta  \simeq 0.01$ and $\Delta m^2\simeq 1.3\times 10^{-8}{\rm
eV}^2$.
\endpage
\sequentialequations
\Ref\DA{R.~Davis Jr.,
 {\it in Frontiers of Neutrino Astrophysics}, eds. Y.~Suzuki
\nextline
 and K.~Nakamura (University Academy Press, Tokyo, 1993);
\nextline
{\it in Proceedings of Second International
 Workshop on Neutrino Telescopes},
\nextline
 eds. M.~Baldo~Ceolin
   (Venezia, 1990) p1.}
\Ref\HI{K.S.~Hirata \journal Phys. Rev. Lett. &63 (89)16;
    \journal ibid &65 (90)1297;
   \nextline
   Y.~Suzuki, {\it in Frontiers of Neutrino Astrophysics}, eds.
   Y.~Suzuki and
   \nextline
   K.~Nakamura (University Academy Press, Tokyo, 1993).}
\Ref\GALLEX{P.~Anselmann \journal Phys. Rev. Lett. &B285 (92)376;390;
     {\bf B314} (93)445.}
\Ref\SAGE{SAGE,
 {\it in Frontiers of Neutrino Astrophysics}, eds. Y.~Suzuki and
\nextline
 K.~Nakamura (University Academy Press, Tokyo, 1993)}
\Ref\BU{J.N.~Bahcall and R.K.~Ulrich
\journal Rev. Mod. Phys. &60 (88)297.}
\Ref\BP{J.N.~Bahcall and M.H.~Pinsonneault
\journal Rev. Mod. Phys. &64 (92)885.}
\Ref\TC{S.~Turck-Chieze et al. \journal Ap. J. &335 (88)415;\nextline
 S.~Turck-Chieze and I.~Lopes \journal Ap. J. &408 (93)347;\nextline
 S.~Turck-Chieze et al. \journal Phys. Rep. &230 (93)57.}
\Ref\MSW{L.~Wolfenstein \journal Phys. Rev. &D17 (78)2369;\nextline
       S.P.~Mikheyev and A.Y.~Smirnov \journal Sov.J.Nucl Phys.
     &42 (85)913}
\Ref\BFP{J.N..~Bahcall, G.B.~Field and W.H.~Press \journal Ap. J.
     &320 (87)L69;\nextline
       B.W.~Filippone and P.~Vogel \journal Phys. Lett.
     &B246 (90)546.}
\Ref\KKT{T.~Kubota, T.~Kurimoto, M.~Ogura and E.~Takasugi \journal
    Phys. Lett. &B292 (92)195;\nextline
    T.~Kubota, T.~Kurimoto and E.~Takasugi
     \journal Phys. Rev. &D49 (93) 2462.}
\Ref\YOS{ H.~Yoshimura \journal Astoro. Phys.
     &178 (72)863.;
    {\it ibid} {\bf 29} (1975)467.;
    \nextline
    {\it ibid} {\bf 52} (1983)363.}
\Ref\SPR{H.~Spruit \journal Solar Phys. &34 (74) 277.}
\Ref\MBV{ M.~B.~Voloshin,M.~I.~Vystotsky and L.~B~Okun
      \journal Sov. Phys. JETP &64 (87)446.}
\Ref\VWET{J.~Vidal and J.~Wudka \journal Phys. Lett. &B249 (90)473;
    \nextline
C.~Aneziris and J.~Schechter \journal Intr. Journal Mod. Phys.
&A6 (91)2375; \nextline
S.~Toshev \journal Phys. Lett.&B271 (91)179; \nextline
A.~Yu.~Smirnov {\it in Proc. Lepton Photon and European
      High Energy Physics Conf.} \nextline(Geneva,1991);\nextline
 E.Kh.~Akhmedov, P.I.~Krastev and A.~Yu.~Smirnov
  \journal Z. Phys. &C52 (91)701.}
\Ref\SMI{A.~Yu.~Smirnov \journal Phys. Lett. &B260 (91) 161.}
\Ref\APS{E.Kh.~Akhmedov, P.I.~Petcov and A.~Yu.~Smirnov \journal
  Phys. Rev. &D48 (93) 2167.}
\Ref\LM{C.S.~Lim and W.J.~Marciano \journal Phys. Rev.
  &D37 (88) 1368.}
\chapter{Introduction}
Three kinds of data of the solar neutrinos flux
are now available from Homestake[1], Kamiokande II and III[2],
GALLEX[3] and SAGE[4] experiments.
These data have shown the deficit of neutrino flux with respect to the
prediction of the standard solar models[5],[6],[7]. The different
deficit rates with respect to detectors show the neutrino energy
dependence of the observed flux. If we take all of these data
seriously, there needs new physics. The MSW
(Mikheyev-Smirnov-Wolfenstein) matter mixing is the most natural
scenario[8] of new physics.

Another characteristic feature is that the Homestake experiment
seem to see the anti-correlation between the number of
sunspots and the counting rates. This anti-correlation
was not observed by the Kamiokande II and III experiments.
Various authors[9] have examined this and concluded that
the anti-correlation was not significant. Recently, we
proposed a new interpretation of the time profiles of the
Homestake and Kamiokande II data[10]. We take
the complex time dependence of the Homestake data
as a real phenomena and consider that the data show
both the short (seasonal) and long (11 years) time variations.
We interpreted the short time variation is originated from the
effect of the twist
of toroidal magnetic fields and the long time variation (11 years)
is due to the change of their magnitudes. The Kamiokande II data are
the averaged ones for about one year so that no short time variation
shows up.

We constructed a simple model of the
twisting toroidal magnetic fields following the simulation by
Yoshimura[11] and made the numerical analysis. We successfully
reproduced the Homestake data as well as the Kamiokande II data[10].
Since we are intended to see purely the effect of the twisting
toroidal magnetic fields, we did not take into account of the
matter mixing effect. Thus our predictions did not have the
neutrino energy dependence. As a result, the GALLEX and SAGE data
are not be explained.

In this paper, we consider the model which have both mechanisms;
the matter mixing and the helicity oscillation in twisting toroidal
magnetic fields. Our concern is to find
a new kind of solution which explain both the deficit of the neutrino
flux (the time averaged profile and the neutrino energy dependence of
the flux) and  the short and long time
variations of the flux. We made the numerical analysis and
examined the allowed region in ($\Delta m^2, \sin^22\theta $) plain.
We found that if  $\Delta m^2\sim 10^{-8}$eV and
$\sin^22\theta \sim 0.01$, all data are reproduced well.

In Sec.2, we discuss the evolution equation and its general features.
A model of the twisting toroidal magnetic fields are briefly explained
in Sec.3 and  the general features of oscillation of this scheme
is discussed in Sec.4. Numerical analysis and the allowed region are
given in Sec.5. Sec.6 is devoted to the discussions.
\chapter{The evolution equation}
Our model consists of two Majorana neutrinos, $\nu _e$ and $\nu _\mu $
which have the transition magnetic moment $\mu$. In the following,
we use $\nu _e$ and $\nu _\mu $ for left-handed neutrinos and
$\bar \nu _e$ and $\bar \nu _\mu $ for their anti-neutrinos, respectively.
The evolution equation for neutrinos which fly along $z$ axis
under magnetic fields is given by
$$i{{d\psi }\over{dz}}=
    \pmatrix{
V_{e} &  s_2\delta & 0 &-\mu B_Te^{i\phi }\cr
s_2\delta  & -V_{\mu }+2c_2\delta  & \mu B_Te^{i\phi } & 0 \cr
   0 & \mu B_Te^{-i\phi }&-V_{e}& s_2\delta \cr
-\mu B_Te^{-i\phi } &0 & s_2\delta   &V_{\mu }+2c_2\delta \cr
}\psi ,
\eqno\eq
$$
where $\psi ^T=(\nu _{e},\nu _{\mu },\bar\nu _{e},\bar \nu _{\mu })$,
$s_2=\sin 2\theta$ and $c_2=\cos 2\theta$ with $\theta$ being a
mixing angle, $\delta =\Delta m^{2}/4E$ with $E$ being
the energy of neutrino and $\Delta m^2$ being the squared mass difference.
The magnetic fields relevant to the helicity oscillation is
the transverse component which are parametrized
by $B_Te^{i\phi }\equiv B_x+iB_y$. If the twist of toroidal magnetic
fields exists, $\phi$ has the $z$ dependence.
The matter potentials for $\nu _e$ and $\nu _\mu $
are denoted by $V_{e}=G_F(2n_{e}-n_{n})/\sqrt{2}$
and $V_{\mu}=G_Fn_{n}/\sqrt{2}$ where $G_F$ is the Fermi constant,
$n_{e}$ and $n_{n}$ are number densities of
electrons and neutrons in the sun, respectively.
For them, we take[12]
 $$\eqalign{V_e=&
    0.195R{^{-1}_\odot} \exp[10.82\sqrt{1-z/ R_{\odot}}]\cr
   V_\mu =&
   0.018R{^{-1}_\odot} \exp[10.82\sqrt{1-z/ R_{\odot}}].\cr
   }
\eqno\eq$$
This is the extension of the OVV (Okun, Voloshin,Vysotsky) model[13].
The model without the matter mixing ($\theta =0$ case) was discussed
by various authors[14],[15],[10] by considering various types of
twisting magnetic fields. The application of this model to
the solar neutrino problem with a realistic model of twisting
magnetic fields was made in Ref.[10].

The importance of the phase $\phi$ can be understood by changing
phases[14],  $\psi ^T\rightarrow $
$\tilde \psi ^T=(e^{i\phi /2}\nu _{e},e^{i\phi /2}\nu _{\mu },
e^{-i\phi /2}\bar \nu _{e},e^{-i\phi /2}\bar \nu _{\mu })$. Then,
apart from the overall phase, $\tilde \psi $ obeys the following  equation
$$i{{d\tilde \psi }\over{dz}}=
    \pmatrix{
V_{e} &  s_2\delta & 0 &-\mu B_T\cr
s_2\delta  & -V_{\mu }+2c_2\delta  & -\mu B_T & 0\cr
0& \mu B_T&-V_{e}-\phi ' & s_2\delta \cr
-\mu B_T & 0&s_2\delta  &V_{\mu }+2c_2 \delta  -\phi '\cr
}\tilde\psi ,
\eqno\eq
$$
where $\phi '=d\phi /dz$. The important role of the phase $\phi$ lies in the
fact that the variation of it works as a potential. Due to this,
various kinds of resonance oscillations will occur
as discussed by Akhmedov, Petcov and Smirnov[16], depending on sizes of
parameters, $c_2\delta $, $s_2\delta $, $\mu B_T$ and $\phi '$.
There are essentially two resonance oscillations of $\nu _e$.
In the following, we assume $\Delta m^2>0$ and $\cos 2\theta >0$ because we are
interested in the transition of $\nu _e$.
\par
\noindent
(i) The MSW matter (flavor) oscillation\par
If $\mu B_T$ is small in comparison with $s_2\delta $,
the MSW matter oscillation occurs.
The resonance condition for $\nu _{e}\leftrightarrow \nu _{\mu }$ oscillation
is
$$V_e+V_\mu  =(\Delta m^2/2E)\cos 2\theta .
\eqno\eq
$$
When the adiabatic condition at the resonance point
$$\left|{1\over{(V_e+V_\mu )}}{{d(V_e+V_\mu )}\over dz}\right| \ll {{\Delta
m^2}\over
{2E}}{{\sin ^22\theta }\over{\cos 2\theta }}
\eqno\eq
$$
is satisfied, $\nu _{e}$ converts fully to $\nu _{\mu }$. If
this condition is not respected, the conversion is not full and
the transition rate is estimated by the numerical computation or
the Landau-Zener formula.
The $\bar \nu _{e}\leftrightarrow \bar\nu _{\mu }$
oscillation does not occur because the condition
$V_e+V_\mu  =-(\Delta m^2/2E)\cos 2\theta $ is not met.
\par
\noindent
(ii) The OVV helicity oscillation in the twisting magnetic fields
\par
If $s_2\delta \equiv (\Delta m^2/4E)\sin 2\theta $ is smaller than $\mu B_T$,
the OVV oscillation in the twisting toroidal magnetic fields
occurs. In this case, the oscillation between two different
flavor neutrinos occurs because Majorana neutrinos can have only
transition moments. The resonance point is determined by
$$\eqalign{V_e-V_\mu  =&(\Delta m^2/2E)\cos 2\theta - \phi '
        \qquad (\nu _{e}\leftrightarrow \bar\nu _{\mu }),\cr
          V_e-V_\mu  =&-(\Delta m^2/2E)\cos 2\theta  -  \phi '
        \qquad (\bar \nu _{e} \leftrightarrow \nu _{\mu }).\cr}
\eqno\eq
$$
We consider the survival probability $P$ of $\nu _{e}$ after passing the
distance $\Delta z$. The rate is expressed by[17]
$$P(\nu _{e}\rightarrow \nu _{e};\Delta z)=1-
B_f\sin ^2{\sqrt{(V_e-V_\mu -2c_2\delta +\phi ')^2 +4(\mu B_T)^2}
\over 2}\Delta z,
\eqno\eq
$$
where $B_f$ is the blocking factor defined by
$$B_f={{4(\mu B_T)^2}\over
{(V_e-V_\mu -2c_2\delta +\phi ')^2 +4(\mu B_T)^2}}.
\eqno\eq
$$
The above formula is valid only when the variation of
$V_e-V_\mu +\phi '$ is small in comparison with
$\mu B_T$. In this formula, the matter potential acts as a blocking
factor. At the resonance point, the blocking factor
disappears ($B_f=1$) and the conversion occurs. The adiabatic
condition
$$
\left|{{d(V_e-V_\mu +\phi ')}\over dz}\right| \ll 4(\mu B_T)^2
\eqno\eq
$$
at the resonance point.
\par
Since the energies of neutrinos spread from about 0.33MeV
($pp$ neutrino) to about 10MeV (${}^8{\rm B}$ neutrino),
$\delta \equiv \Delta m^2/4E$
varies about factor 30 so that neutrinos may
receive different types of effects.
We look for the situation where the $pp$
neutrinos receive mainly the MSW
matter oscillation effect, while the ${}^8{\rm B}$
neutrinos receive the OVV helicity oscillation effect in the
twisting toroidal magnetic fields. In this situation, the time
variations occur for ${}^8{\rm B}$ neutrinos, but not for
$pp$ neutrinos. This possibility is realized when $\delta$ and
$\mu B_T$ are the same size.
\par
\chapter{A model of twisting toroidal magnetic fields}
Here we give a brief summary of the model of twisting toroidal
magnetic fields which is proposed in our previous papers[10].
As in Fig.2 in Ref.10, We assume two tori in the
convective area of the sun; one in the northern hemisphere and
the other in the southern hemisphere  parallel to the equator.
We assume that toroidal magnetic fields locate in the torus
and toroidal magnetic fields twist along it.
This twist is parametrized by a parameter $X$,
the distance along the torus to wind once. As in Fig.5 in Ref.10,
we parametrize the configuration of the
torus in the sun's cross section.
The torus is parametrized by the latitude of its center $\Delta$,
its radius $a$, the distance between the center of torus and the
center of the sun $b$.
\par
The latitude of the neutrino path is parametrized by $\lambda$ which
is between  $-7.25^\circ$ (the southern hemisphere) and
$7.25^\circ$ (the northern hemisphere). Since $\lambda$ is small,
neutrinos pass through almost around the edge of the toroidal
magnetic fields so that we assume that the strength of magnetic fields
is constant along the neutrino path. We specify the
position of the entrance $z_0$ of a neutrino to
magnetic fields and the departure $z_1$ from it.
They are given by
   $$z_{0,1} \equiv b\cos(\Delta -\lambda )\pm \sqrt{a^2-b^2\sin^{2}
             (\Delta -\lambda )}.\eqno\eq$$
\par
The twist of troidal magnetic fields is  generated by the
differential rotation of the sun and the global convection of the
plasma fluid in the convective area. By solving the dynamo equation,
Yoshimura[11]
showed that the development of the twist of the troidal magnetic
fields is the origin of the cyclic oscillations of polarity reversals
in every 11yr. The direction of the twist
depends on whether magnetic fields lie in the northern or the
southern hemisphere. The global structure of the twist is determined
by the Coliori's force which acts on the global convection.
By comparing the simulation by Yoshimura with our simple model,
we found that the variation of the phase $\phi '$ is given by[10]
$$\phi '(z)= -{\rm sign}(\lambda )
  {{2\pi /X}\over{1+(2\pi /X)^2 [b \cos (\Delta -\lambda )-z]^2}}.
\eqno\eq$$
The characteristic feature is that
$\phi '<0$ if neutrinos pass through the northern
hemisphere and $\phi '>0$ if they do the southern hemisphere. This sign
difference will give the important affect on the deficit rate of the
neutrino flux[10].
Various parameters defining the
toroidal magnetic fields are fixed by comparing our model with
the simulation by Yoshimura,
$\Delta =15^{\circ}$, $a=0.1694R_\odot$ and $b=0.8813R_\odot$.
\par
\chapter{General features of oscillation}
We see some qualitative features of the neutrino survival rate.
In order to make some concrete arguments, we take $\sin 2\theta =0.1$,
$(\mu B_T)\sim 4\times 10^{-10}\mu _B$kG$\sim 8/R_\odot$
and $X=\pi R_\odot$ (about two turns)
so that $\phi '\sim \pm 2/R_\odot$.
Also, we restrict $\lambda$ in two cases
$\lambda =7^\circ$ (northern hemisphere, around
September) and $\lambda =-7^\circ$ (southern hemisphere, around April).
With these values, we obtain  $z_0=0.758R_\odot$
and $z_1=0.991R_\odot$.

The survival rate is given as a function of $y$ which is defined
as
$$y \equiv {(E/1{\rm MeV})\over(\Delta m^2/1{\rm eV}^2)}
   ={{3.5\times 10^9}\over R_\odot}{E\over{\Delta m^2/(1{\rm eV}^2)}}.
\eqno\eq
$$
The evolution equation is solved from the center to the
surface of the sun with the initial condition $\psi =(1,0,0,0)$ at
the center of the sun where $\nu _{e}$ is created.
In
\Fig{Energy dependence of the survival probability of $\nu _e$.
We estimated it for $\sin 2\theta =0.1$,
$(\mu B_T)\sim 4\times 10^{-10}\mu _B$kG and $X=\pi R_\odot$.
The solid (dashed) line corresponds to the case where neutrinos pass
through the northern (southern) hemisphere. The MSW matter oscillation
mechanism dominates when
$y \equiv (E/1{\rm MeV})/(\Delta m^2/1{\rm eV}^2)$
is smaller than $3\times 10^7$ so that no seasonal difference
arises. When $y>3\times 10^7$, seasonal differences appear due to
the twist of toroidal magnetic fields},
we show our result about the
$y$ dependence of the survival probability
$P(\nu _{e}\rightarrow \nu _{e})$ observed on earth. The solid (dotted)
line represents the survival probability of $\nu _e$ when
it passes through the northern (southern) hemisphere. The seasonal
difference arises for $y > 2\times 10^7$ where the effect of the twisting
magnetic fields becomes dominant.
In
\Fig{ Energy dependence of neutrino components. We show where
$\nu _e$ transforms when neutrinos pass through the northern hemisphere
(a) and the southern hemisphere (b). For $y<3\times 10^7$, $\nu _e$
transforms mainly to $\nu _\mu $, while for $y>3\times 10^7$, $\nu _e$
does to $\bar \nu _\mu $.},
we show  to which
spices  $\nu _{e}$ transformed. Fig.2b(c) corresponds to
the case where neutrinos pass through northern (southern) hemisphere.
$\nu _e$ transforms mainly to $\nu _\mu $ up to $y \sim  3\times 10^7$. When
$y$
is greater than this value, $\nu _e$ transforms to $\bar \nu _\mu $.
The results in these figures may be understood qualitatively
as follows:
\par
\noindent
(i) $y<2\times 10^5$\par
The resonance condition of the
flavor oscillation in Eq.(4) is not satisfied and the
resonance oscillation does not occur.
\par
\noindent
(ii) $2\times 10^5 < y < 10^6$\par
In this region, the resonance point is inside the sun and
the adiabatic condition is satisfied. Thus, $\nu _{e}$ transforms
fully to $\nu _{\mu }$. The resonance condition of helicity
oscillation may be satisfied, but the resonance point
is no in magnetic fields which are located in the region
$0.758R_\odot<z<0.991R_\odot$.  Thus, the helicity conversion
does not occur.
\par
\noindent
(iii) $10^6<y<3\times 10^7$ \par
In this region, the non-adiabatic flavor transition occurs.
As in Figs.1b and 1c, a part of $\nu _{e}$ transforms into
$\nu _{\mu }$.
\par
\noindent
(iv) $3\times 10^7<y<5\times 10^8$\par
Since $\mu B_T$ is assumed to be  around $8/R_\odot$, it
is larger than $(\Delta m^2/4E)\sin 2 \theta $. Thus
the helicity transition $\nu _{e}\rightarrow \bar \nu _{\mu }$ occurs.
The resonance condition is expressed by
$Y_E=0.177\exp(10.82\sqrt{1-z/R_\odot})$ and
the adiabatic condition in Eq.(9) is expressed by
$$\left|{Y_E\over{\log |Y_E|}}\right|\ll 1,
\eqno\eq
$$
where $Y_E=1.8\times 10^9/y-\phi 'R_\odot$.
In order for the adiabatic condition is satisfied,
$|Y_E| \ll 1$ is required.
This means that the resonance point should be
very close to the surface and thus the adiabatic condition is not
satisfied. In this non-adiabatic case, the estimation of the
transition probability is rather complicated.
We see from Fig.1a that in this $y$ region the seasonal difference
appears due to the effect of the twist $\phi '$.
The role of $\phi '$ is to shift the resonance point. When neutrinos
pass through the northern (southern) hemisphere, $\phi '$ is negative
(positive) and the resonance point $z_{\rm r}$ is smaller (larger)
than the non-twisting case. For example, for $y\sim 10^8$,
the resonance point
determined from Eq.(7) is $z_{\rm r}\sim 0.81R_\odot$ for the northern
hemisphere and  $z_{\rm r}\sim 0.83R_\odot$ for the southern hemisphere.
Since the transition is non-adiabatic, the transition develops
gradually until neutrino reaches to the surface. It turns out
that for $y\sim 10^8$  and $\phi '<0$ (northern hemisphere),
the survival rate becomes minimum before neutrino reaches
to the surface so it oscillates back at the surface. On the other hand,
for $\phi '>0$ (southern hemisphere) the minimum value is achieved
at the surface. This is the reason why the seasonal difference
appears.
\par
\noindent
(v) $10^9>y>5\times 10^8$\par
In this region, the pure helicity transition in the matter
discussed in Ref.10 is realized. Mostly, $\nu _{e}$
transforms to $\bar \nu _{\mu }$. The general tendency is that the
suppression of $\nu _{e}$ flux is strengthen when neutrinos
pass through the northern hemisphere ($\phi '<0$) and is weaken when
neutrinos pass through the southern hemisphere ($\phi '>0$).
The transition rate depends sensitively on the strength of
magnetic fields as shown in Ref.10.
\par
Since the average energy of $pp$ neutrinos is 0.33MeV and that of
${}^8{\rm B}$ neutrinos is 10MeV, there is about 30 times difference.
The experimental data by Homestake, Kamiokande II and III,
SAGE and GALLEX suggest that $pp$ neutrinos are less suppressed than
${}^7{\rm Be}$ and ${}^8{\rm B}$  neutrinos. From Fig.1a,
there seem two possible solutions:
One is the region $y\sim (10^6\sim 10^7)$ where the MSW matter oscillation
works.  The $pp$ neutrinos are less suppressed than ${}^7{\rm Be}$
and ${}^8{\rm B}$ neutrinos and no time dependence of fluxes appears.
In this solution, $\nu _{e}$ transforms to $\nu _{\mu }$.
The other is the region $y\sim (10^8\sim 10^9)$ where
the $pp$ neutrinos are less suppressed and have small time dependence.
On the other hand, the ${}^7{\rm Be}$ and ${}^8{\rm B}$ neutrinos are
more suppressed and two kinds of time dependencies, long (11yr)
and shot (seasonal) oscillation appear. This is the region where
we are seeking.
\par
\chapter{A new solution}
In order to solve the evolution equation, we have to give
the size of the twist $\phi '$ and the time variation of the
strength of magnetic fields. Although the twist should exist,
the size of $\phi '$ is an unknown parameter. Here we take
$X=\pi R_\odot$. We guess the time variation of magnetic fields
from the variation of the sun spot number,
$B_T\propto
\sqrt{{\rm sun} {\rm spot} {\rm number}}$. We also assumed the maximum
strength $(\mu B_T)_{\rm max}=10.5\times 10^{-10}\mu _B$kG.
With these choices of values,
we can estimate the neutrino flux at each year and each season.
For the latitude of the neutrino path, we consider
$\lambda =7^{\circ}$ (around September) and $\lambda =-7^{\circ}$ (around
April).
According to the simulation by Yoshimura, magnetic fields
on neutrino paths
from August to October are similar to those of September and thus
the calculation with $\lambda =7^{\circ}$ will valid for the period
from August to October.
Similarly, the calculation with $\lambda =-7^{\circ}$ will be valid for
the period from March to May.
\par
We now seek a solution in the region $y\sim 10^8$ following the discussion
given in the previous section. We considered three
experiments, Homestake, Kamiokande II and III and GALLEX.
For Kamiokande II and III,
we used the formula
$$
P({\rm KII})=P({}^8{\rm B})+0.11\bar P({}^8{\rm B}), \eqno\eq$$
where $P$ and $\bar P$ represent the probabilities of $\nu _{e}$ and
$\bar \nu _{\mu }$, respectively. The symbol ${}^8{\rm B}$ in the
parenthesis show the survival rate of ${}^8{\rm B}$ neutrinos
which is evaluated with use of the  average  energy
$E_\nu =$10MeV.  The contribution
from the anti-neutrino arises because Kamiokande II and III detectors
are sensitive to it also. GALLEX and Homestake detectors
can catch other than ${}^8{\rm B}$ neutrinos. We used the formula
$$\eqalign{P({\rm GALLEX})=&{70.8\over 132}P(pp)
                +{3.0\over 132}P(pep)
                +{34.3\over 132}P({}^7{\rm Be})\cr
        &\qquad +{14.0\over 132}P({}^8{\rm B})
  + {3.8\over 132}P({}^{13}{\rm N})+{6.1\over 132}P({}^{15}{\rm O})},
\eqno\eq
$$
$$\eqalign{P({\rm Homestake})=&{0.2\over 7.9}P(pep)
                +{1.1\over 7.9}P({}^7{\rm Be})\cr
        &\qquad +{6.1\over 7.9}P({}^8{\rm B})
  + {0.1\over 7.9}P({}^{13}{\rm N})+{0.3\over 7.9}P({}^{15}{\rm O})}.
\eqno\eq
$$
For $pp$, ${}^{13}{\rm N}$ and ${}^{15}{\rm O}$ neutrinos, the
survival rates are estimated by using their average energies
0.33, 1.1 and 1.44MeV.
\par
For Kamiokande II and III and GALLEX data, we demand that the time
averages of our calculated rates
agree with their averages within 1 standard deviation. We relaxed this
constraint for the Homestake data because the data points
changes with time so that the time average may not give
a good measure for agreement. We put the importance on
the time variation more than the average value for this case.
We require that the time average of our
estimates agree with the average experimental value
within four standard deviation.

Allowed regions are shown in
\Fig{Allowed domains by three experiments. Homestake data restrict
the area to the domains surrounded by the solid lines; a large
area in the upper-right corner and many small islands.  Kamiokande
II and III data restrict to the areas surrounded by the dotted lines;
relatively large areas with $\Delta m^2\sim 10^{-8}{\rm eV}^2$
and $\sim (2\sim 3)10^{-9}{\rm eV}^2$ and several islands. GALLEX data
restrict to the large area from $\Delta m^2\sim 3\times 10^-7$
to $4\times 10^8{\rm eV}^2$ which is surrounded by the dash-dotted lines.
The intersection of these areas is the small area with the shade.}.
The area covered by dash-doted lines,
solid lines and dotted lined
are the allowed regions by GALLEX data, Homestake data and
Kamiokande II and III data, respectively. We found an unique
shaded area which satisfies three experimental restrictions.
The mixing angles and the squared mass in
the allowed domain are $\sin ^2 2\theta \sim 1\times 10^{-2}$ and
$\Delta m^2\sim  1.3\times 10^{-8}{\rm eV}^2$.
\par
We show the comparison between our predictions and the data
in
\Fig{The comparison of our predictions with three data.
With the choice of some values of $\Delta m^2$ and $\sin 2\theta $
in the shaded area, we estimated
the neutrino yields for three experiments. The comparison of
our prediction with Homestake data is shown in (a), with
Kamiokande II and III data in (b) and with GALLEX data in (c).}.
Fig.4a show the comparison with Homestake data, Fig.4b with
Kamiokande II and III data and Fig.4c with GALLEX data.
Our predictions for the Homestake data show
the short (seasonal) and long (11yr) time variations and reproduce
the data pretty well.
We also predict these time variations for the
Kamiokande II and III data, but the comparison with respect to
the short time variation will be premature at present
because the Kamiokande II and III are given as the average for
about one year. There appears no seasonal time variation if
one year average data are taken.
As for the GALLEX data, the main component
is $pp$ neutrinos. Thus we predict almost no time
variation for GALLEX data as seen in Fig.4c.
The agreements of our predictions and three kinds of data seem
to be pretty good.
\chapter{Discussions}
We presented a new solution to explain the solar neutrino problems,
the missing flux problem and the short and long time variations.
We used a simple model of twisting magnetic fields in the sun
and showed that this type of magnetic fields
can give a new kind of solution for the solar neutrino problem.
Our model predicts the short and long time variations which should
appear in the Homestake and Kamiokande II and III data, and weak
time dependence for GALLEX and SAGE data. Our predictions
seems to reproduce these data pretty well.
We also estimated to which type of neutrinos  $\nu _e$ converts.
{}From Figs.1b and 1c, the $pp$ neutrino converts mainly to $\nu _\mu $,
while the ${}^7{\rm Be}$ and ${}^8{\rm B}$ neutrinos do to $\bar \nu _\mu $.
Thus, the situation in our model is more complicated than the
MSW matter oscillation case. In order to clarify these situations,
we have to wait the future experiments.

\endpage
\refout
\endpage
\figout
\end